\documentclass[twocolumn,showpacs,preprintnumbers,amsmath,amssymb,superscriptaddress]{revtex4}
\usepackage{graphicx}
\usepackage{dcolumn}
\usepackage{bm}

\begin{document}
\title{Determining the accuracy of  spatial
gradient sensing using statistical mechanics}
\author{Bo Hu, Wen Chen, Wouter-Jan Rappel and Herbert Levine}
\address{ Department of Physics and Center for Theoretical Biological
Physics, University of California, San Diego, La Jolla, CA
92093-0374 }
\date{\today}

\begin{abstract}

Many eukaryotic cells are able to sense chemical
gradients by directly measuring spatial concentration differences.
The precision of such gradient sensing is limited by fluctuations
in the binding of diffusing particles to
specific receptors on the cell surface. Here, we explore the
physical limits of the spatial sensing mechanism by modeling the
chemotactic cell as an Ising spin chain subject to a spatially
varying field. This allows us to derive the \emph{maximum likelihood
estimators} of the gradient parameters as well as explicit
expressions for their asymptotic uncertainties. The accuracy 
increases with the cell's size and our results demonstrate that 
this accuracy be further increased 
by introducing a non-zero cooperativity between neighboring 
receptors. 
Thus, consistent with recent experimental data, it 
is possible for small bacteria to perform spatial measurements of
 gradients.
\end{abstract}

 \maketitle

Cells often direct their motion under the guidance of chemical
gradients. This is essential for critical biological functions
including neuronal development, wound repair and cancer spreading
\cite{Parent,Haastert}. To detect gradients, small organisms like
bacterial cells usually employ a temporal sensing strategy by
measuring and comparing concentration signals over time along their
swimming tracks \cite{Macnab, Segall, Bray, Sourjik}. In contrast,
eukaryotic cells are sufficiently large to implement a spatial
sensing mechanism, as they can measure the concentration differences
across their cell bodies. Measurements for both strategies are
accomplished by specific cell-surface receptors which diffusing
chemical particles (ligands) can bind to. Spatial sensing among
eukaryotes exhibits a remarkable sensitivity to gradients of merely
1-2\% across the cell \cite{Song, HaastertBJ, Fuller}. Given the
dynamic fluctuations in ligand-receptor interaction, the receptor
signal is inherently noisy, as demonstrated by single-cell
imaging experiments \cite{Ueda1, Ueda2}. This naturally raises a
question concerning the reliability of spatial gradient sensing.

In analyzing bacterial chemotaxis, Berg and Purcell showed that the
minimal uncertainty of mean concentration measurements is set by the
diffusion of ligand particles \cite{Berg1977}. This work has been
extended to include ligand-receptor binding effects and possible
receptor cooperativity
\cite{Bialek2005,Wang1,Bialek2008,EndresPNAS,EndresPRL}. The spatial
sensing program concerns the acquisition of information regarding
the asymmetry in space (the gradient steepness and direction). The
accuracy of gradient measurements should similarly be limited by
physical laws governing diffusion and stochastic ligand-receptor
dynamics. There have been some studies on the limits to spatial sensing,
but either for idealized sensing mechanisms that ignore 
the receptor kinetics \cite{EndresPNAS} or
based on specific transduction models \cite{Wouter}. In this Letter,
we address this problem in a general way using a statistical
mechanical approach where we view the surface receptors as a
(possibly coupled) spin chain and treat the chemical gradient as a
perturbation field. By calculating the system's partition function,
we are able to derive the physical limits of gradient sensing for 
both independent receptors and for receptors exhibiting cooperativity. 
These limits allows us to predict that the strategy of
spatial sensing may not be exclusive to large eukaryotic cells but may
also be applicable to some bacterial cells \cite{Thar}, especially
with the aid of receptor cooperativity.

\begin{figure}
\scalebox{0.35}[0.32]{\includegraphics{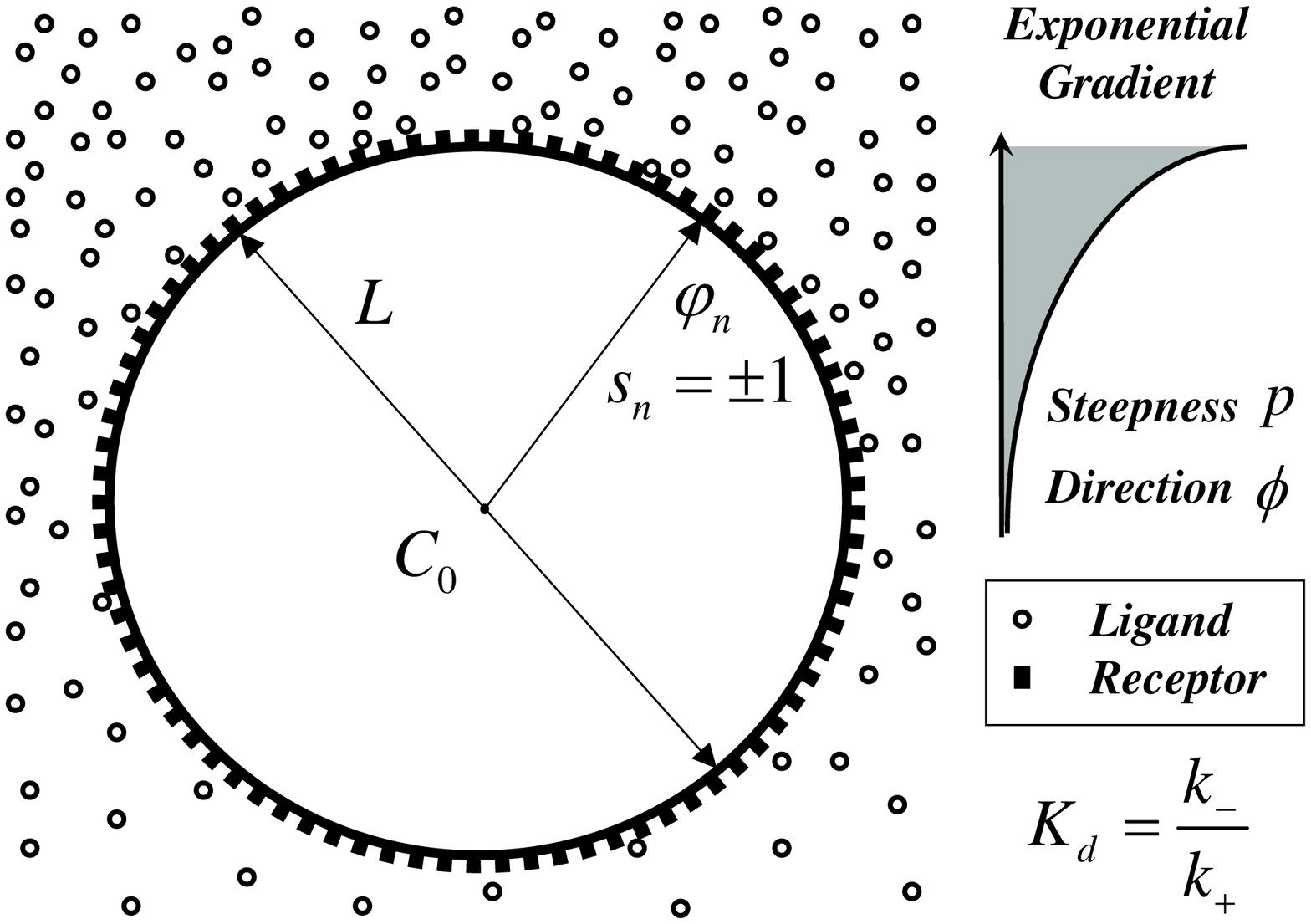}} \caption{Schematic
representation of our model: a circular cell, covered with 
receptors, is placed in an 
exponential gradient. The forward and backward rates
$k_{\pm}$ control the transition between the bound and unbound
states for the receptors.
}
\end{figure}

We consider a circular cell with diameter $L$ immersed
in a chemoattractant gradient (Fig. 1). We suppose that there are
$N$ receptors distributed at equally spaced intervals on the cell's
perimeter \cite{footnote1}. The angular coordinates of these
receptors are indicated by $\varphi_n=2\pi n/N$ for $n=1,...,N$. For
analytical convenience, we assume that the gradient field takes an
exponential profile, as was recently realized in experiments
utilizing the social amoeba {\em Dictyostelium} \cite{Fuller, Hu}.
The local concentration at the n-$th$ receptor is
$C_n=C_0\exp{\left[\frac{p}{2}\cos(\varphi_n-\phi)\right]}$, where
$C_0$ is the background concentration, $p\equiv\frac{L}{C_0}\frac{d
C}{d r}$ denotes the gradient steepness, and
$\phi$ indicates the gradient direction. Like a spin in physics,
each receptor switches between two states: either active ($s_n=+1$)
or inactive ($-1$). For independent receptors, a receptor is
activated only if it is bound by ligand and inactive otherwise. Let
the energy associated with the state $s_n=+1$ (or $-1$) be
$-\varepsilon_n$ (or $+\varepsilon_n$) in units of the thermal
energy $k_BT$. Then the ``on" probability of the n-$th$ spin is
given by the Boltzmann distribution:
$P_{on}=e^{\varepsilon_n}/(e^{\varepsilon_n}+e^{-\varepsilon_n})$.
For simple receptor-ligand kinetics (Fig. 1), we have
$P_{on}=C_n/(C_n+K_d)$ in chemical equilibrium where $K_d=k_-/k_+$
is the dissociation constant. Therefore, the free energy has the
expression:
\begin{equation}
\varepsilon_n=\frac{1}{2}\ln\frac{C_n}{K_d}=\frac{1}{2}\ln\frac{C_0}{K_d}+\frac{p}{4}\cos(\varphi_n-\phi)\equiv
\alpha_0+h_n.
\end{equation}
We define three statistical quantities
$(z_0,z_1,z_2)\equiv\left(\sum_ns_n,\frac{1}{2}\sum_ns_n\cos\varphi_n,
\frac{1}{2}\sum_ns_n\sin\varphi_n\right)$ where 
$z_0$ is a measure of the average receptor activity and 
where $z_1$ and $z_2$ measure the asymmetry in the receptor activity.
Using the transformation $(\alpha_1, \alpha_2)\equiv(p\cos\phi, p\sin\phi)$ 
we can write the system's Hamiltonian as
$\mathcal{H}_N\{s_n\}=-\sum_n\varepsilon_n
s_n=-\alpha_0z_0-(\alpha_1z_1+\alpha_2z_2)/2$ and compute its
logarithm partition function as follows,
\begin{eqnarray}
\ln\mathcal{Q}_N&=&\ln\prod_{n=1}^N(e^{\varepsilon_n}+e^{-\varepsilon_n})
=\sum_{n=1}^N\ln[2\cosh(\alpha_0+h_n)]\nonumber\\
&=&N\ln(2\cosh{\alpha_0})+\frac{Np^2}{64\cosh^2\alpha_0}+\mathcal{O}(p^4),
\end{eqnarray} where in the last step the summand is expanded in
powers of $p$ and the sum is replaced by an integral over
$[0,2\pi]$.

The partition function contains all the thermodynamic information we
need to infer the gradient parameters $p$ and $\phi$, or
alternatively, the transformed parameters $\alpha_1$ and $\alpha_2$.
Since $p^2=\alpha_1^2+\alpha_2^2$, we have by Eq. (2):
\begin{eqnarray}
\mathrm{E}[z_{1,2}]&=&2\frac{\partial\ln\mathcal{Q}_N}{\partial\alpha_{1,2}}
=\frac{\alpha_{1,2}NC_0K_d}{4(C_0+K_d)^2}+\mathcal{O}(p^3),\\
\mathrm{Var}[z_{1,2}^2]
&=&4\frac{\partial^2\ln\mathcal{Q}_N}{\partial\alpha_{1,2}^2}=\frac{NC_0K_d}{2(C_0+K_d)^2}+\mathcal{O}(p^2).
\end{eqnarray}
In addition, one can check that $\mathrm{Cov}[z_1,z_2]=0$. Thus, for
small $p$, the joint probability density of $z_1$ and $z_2$ is
\[
f(z_{1,2}|\alpha_{1,2})\approx\frac{1}{2\pi\sigma^2}\exp\left[-\frac{(z_1-\mu\alpha_1)^2+(z_2-\mu\alpha_2)^2}{2\sigma^2}\right],
\]
with $\mu\equiv NC_0K_d/(4(C_0+K_d)^2)$ and $\sigma^2=2\mu$
\cite{Hu}. It is easy to show that the \emph{maximum likelihood
estimator} (MLE) \cite{SteveKay} of $\alpha_{1,2}$ is
$\widehat{\alpha}_{1,2}=z_{1,2}/\mu$. As an orthogonal
transformation, the MLE of $p$ and $\phi$ are given by
$\widehat{p}=\sqrt{\widehat{\alpha}_1^2+\widehat{\alpha}_1^2}=\mu^{-1}\sqrt{z_1^2+z_2^2}$
and
$\widehat{\phi}=\arctan(\widehat{\alpha}_2/\widehat{\alpha}_1)=\arctan(z_2/z_1)$,
respectively. By the properties of MLE, both $\widehat{p}$ and
$\widehat{\phi}$ tend to be unbiased and normal in the large $N$
limit, i.e., $\widehat{p}\xrightarrow{d} \mathcal{N}(p,\sigma_p^2)$
and $\widehat{\phi}\xrightarrow{d} \mathcal{N}(\phi,
\sigma_{\phi}^2)$, where ``$\xrightarrow{d}$" denotes convergence in
distribution. The asymptotic variances in the 
gradient steepness and direction, $\sigma_p^2$ and
$\sigma_{\phi}^2$, can be derived from the Fisher information matrix
\cite{SteveKay}, which is diagonal since 
$p$ and $\phi$ are independent. Thus,
\begin{eqnarray}
\sigma_p^2=1/\mathrm{E}\left[(\partial_p
\ln f)^2\right]&=&\frac{\sigma^2}{\mu^2}=\frac{2}{\mu}=\frac{8(C_0+K_d)^2}{NK_dC_0},\\
\sigma_{\phi}^2=1/\mathrm{E}\left[(\partial_{\phi} \ln
f)^2\right]&=&\frac{\sigma^2}{\mu^2
p^2}=\frac{8(C_0+K_d)^2}{Np^2K_dC_0}.
\end{eqnarray}
and thus $\sigma_{\phi}^2=\sigma_p^2/p^2$.
According to the Cram\'{e}r-Rao inequality, $\sigma_p^2$ and
$\sigma_{\phi}^2$ set the lowest uncertainties of gradient
measurements from an instantaneous sampling of the receptor states
\cite{SteveKay}. 
The approximation for both
variances is plotted in Fig. 2 as a function of the two 
parameters characterizing the gradient: the 
background concentration $C_0$ (Fig. 2A) and the gradient 
steepness $p$ (Fig. 2B). 
We have also performed 
Monte-Carlo simulations in which 80000 receptors are 
uniformly distributed along the circular cell membrane. 
Computing our statistical quantities for 5000 
independent realizations, we determined the measurement 
errors and have plotted them as symbols in Fig. 2. 
The analytical results agree well with the numerically obtained 
values. 
From Fig. 2, we can see that  the variances reach a minimum for $C_0=K_d$ while
only the error in the gradient direction 
depends on the steepness of the gradient ($\sigma_{\phi}^2\sim
p^{-2}$).
Thus, since $p=p_0L$ (with
$p_0=\frac{1}{C_0}\frac{dC}{dr}$) increases with the 
cell's size, larger cells are able to sense the gradient direction with
higher accuracy.

\begin{figure}
\scalebox{0.30}[0.30]{\includegraphics{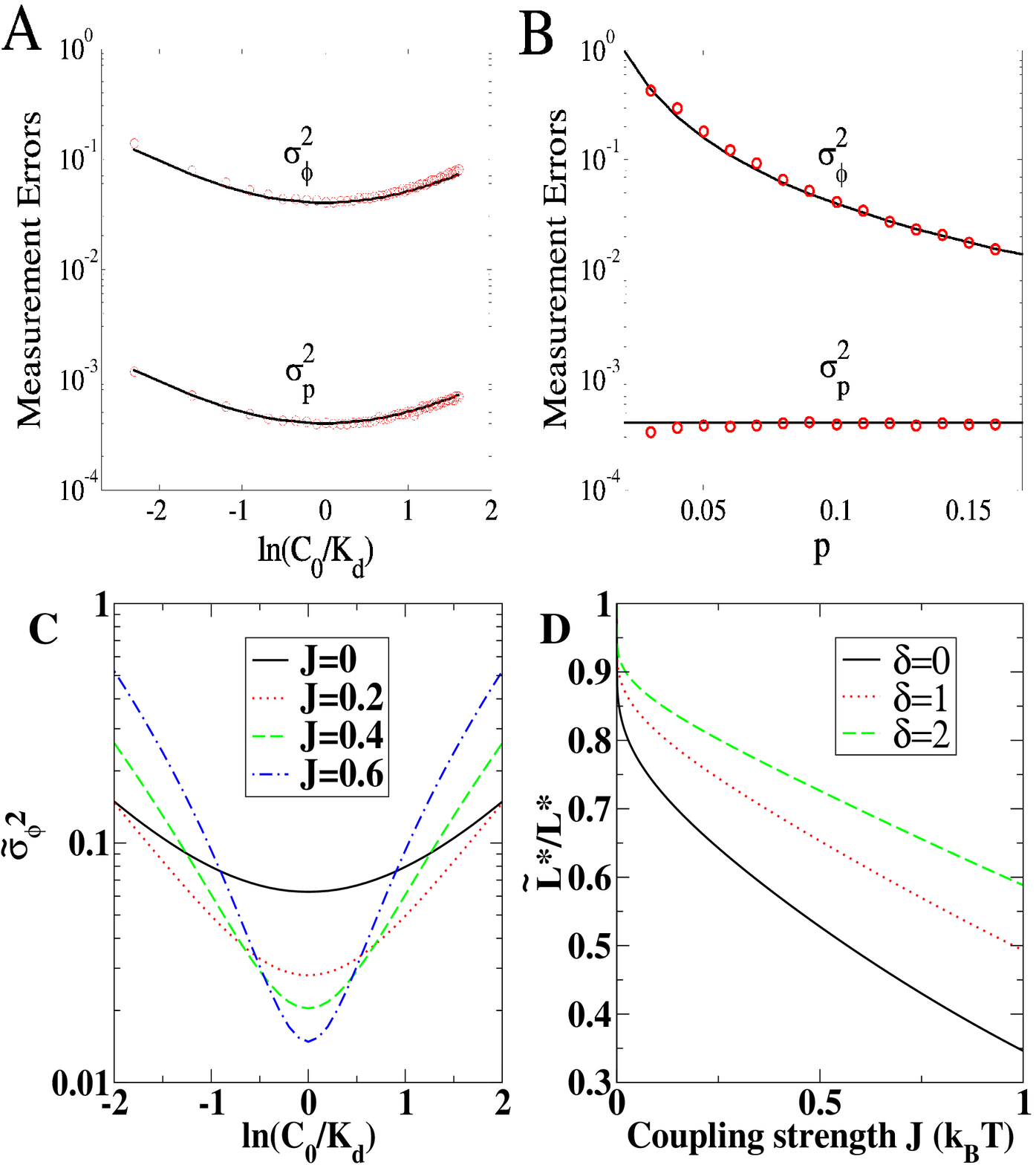}} 
\vspace{-1cm}
\caption{(Color
online). (A) The uncertainties $\sigma_p^2$ and $\sigma_{\phi}^2$
versus $\ln(C_0/K_d)$ (A; $p=10$\%, $N=80000$) and 
versus $p$ (B; $C_0=K_d$, $N=80000$). The solid lines correspond to the 
approximate analytical expressions while the 
symbols are the result of 5000
independent Monte-Carlo realizations.
(C) $\widetilde{\sigma}_{\phi}^2$ as a function of
$\ln(C_0/K_d)$ for
different values of $J$. (D) The critical cell size
below which spatial gradient sensing is ineffective, normalized
by the critical cell size in the absence of
cooperativity, as a function of the cooperativity strength.
In C and D, we have chosen $N=80000$ and $p=8$\%.
}
\end{figure}

The above results are derived from a single snapshot of the system. If the
cell integrates receptor signals over some time interval
$\mathcal{T}$, then averaging over multiple measurements can
appreciably reduce the errors of gradient sensing. However, the
capacity of such averaging is limited by the expected time it takes
for every independent measurement. As shown in \cite{Wang1,
HaastertBJ}, the time to complete a single measurement is roughly
twice the system's correlation time $\tau$ resulting from the diffusion and
binding of ligand molecules, leading to 
a reduction of the variance:
$\sigma_{p,\mathcal{T}}^2=\frac{2\tau}{\mathcal{T}}\sigma_{p}^2$. 
The correlation time is given by
$\tau=\tau_{\mathrm{rec}}+\tau_{\mathrm{diff}}$, where
$\tau_{\mathrm{rec}}=1/(k_-+C_0k_+)$ is the timescale of
receptor-ligand reaction and $\tau_{\mathrm{diff}}$ describes the
diffusive transport time of ligands. 
Let $\eta\equiv\tau_{\mathrm{diff}}/\tau_{\mathrm{rec}}$, 
then the measurement is reaction-limited if $\eta\ll1$ and
diffusion-limited if $\eta\gg1$. 
From the above arguments we find that
averaging signals over $\mathcal{T}$ yields a lower uncertainty of
the gradient estimate,
\begin{eqnarray}
\sigma_{p,\mathcal{T}}^2 \simeq \frac{2\tau}{\mathcal{T}}\sigma_{p}^2
=\frac{4 \tau_{rec}(1+\eta)}{\mu \mathcal{T}}
=\frac{16(1+\eta)}{N\mathcal{T}k_-}\left(1+\frac{K_d}{C_0}\right)
\end{eqnarray}
For small background
concentrations ($C_0\ll K_d$),  
$\tau_{\mathrm{diff}}=N/(2\pi LD K_d)$ where $D$ denotes the
ligand diffusion coefficient \cite{Berg1977, Wang1, Lauffenburger},
and the uncertainty reduces to 
$ \sigma_{p,\mathcal{T}}^2 \simeq
16/(N\mathcal{T}C_0k_+)+8/(\pi \mathcal{T}DLC_0)$.
This expression contains two terms: the first one is
determined by the chemical kinetics, and the second one, up to a
geometric constant, is exactly the Berg-Purcell limit
\cite{Berg1977} or the result recently derived in \cite{EndresPNAS}.
We can derive similar results for
the direction inference, since
$\sigma_{\phi,\mathcal{T}}^2=\sigma_{p,\mathcal{T}}^2/p^2$. 
For typical eukaryotic cells, it has been
estimated  \cite{Wang1, Lauffenburger} that $\eta\ll1$, which implies $\sigma_{\phi,
\mathcal{T}}^2\simeq16(1+K_d/C_0)/(Np^2\mathcal{T}k_-)\sim1/(Np^2)$.
We can relate the uncertainty in the direction measurement to the
cell's size $L$.  Assuming that the 
number of receptors in our model scales with the
cell size as $N=N_0L^\delta$ with $0\leqslant\delta\leqslant2$, we
find $\sigma_{\phi,\mathcal{T}}^2\sim L^{-(2+\delta)}$. 
For comparison, the Berg-Purcell analysis 
considered only an average concentration measurement and  scales as
$\sigma_{c,\mathcal{T}}^2\sim L^{-1}$ \cite{Berg1977}. 
Not surprisingly, our results indicate that spatial directional 
sensing can be more sensitive to the cell's size.

Our analysis above, 
which extends beyond the Berg-Purcell framework by 
providing a direct calculation of the directional sensing limit
$\sigma_{\phi,\mathcal{T}}^2$, 
was carried out  for {\em independent} receptors, as is
assumed to be the case for most eukaryotic cells that have been studied
to date. 
We now ask, what if there is receptor cooperativity as has
been found in many bacterial cells \cite{Shi, Mello, Wingreen}?
Intuitively, short-range interactions make it possible for receptors
to collectively respond and thus sharpens the asymmetry of receptor
signals. It is natural to speculate that such enhanced sensitivity
may set new and lower limits for directional sensing. To incorporate
potential receptor cooperativity, we extend our model to include
a nearest-neighbor interaction $J$ (again, in units of
the thermal energy $k_BT$). 
Now, the activity of a receptor,
again represented by $\{s_n=\pm1\}_{n=1}^N$, 
is determined not only by the local chemical concentration but also by
the states of its neighboring receptors. This means that an unbound
receptor is not necessarily inactive, as it may have been affected
by active, nearby receptors.

Because the local concentration is identical for nearest-neighbor
sites (i.e., $\varepsilon_n=\varepsilon_{n\pm1}$), the Hamiltonian
of our Ising chain can be written in a symmetric form:
$\widetilde{\mathcal{H}}_N\{s_n\}=-\sum_{n=1}^N[Js_ns_{n+1}
+\varepsilon_n(s_n+s_{n+1})/2]$,
with the boundary condition $s_{N+1}=s_1$. The corresponding partition
function is
$\widetilde{\mathcal{Q}}_N=\sum_{s_1}...\sum_{s_N}e^{-(H_0+H_1)}$,
where $H_0\equiv-\sum_n[Js_ns_{n+1}+\alpha_0(s_n+s_{n+1})/2]$
represents the Hamiltonian of an isotropic reference system and
where
$H_1\equiv-\sum_ns_nh_n=-\frac{p}{4}\sum_ns_n\cos(\varphi_n-\phi)$
results from the spatial heterogeneity of the concentration. For
small $p$, one can view $H_1$ as a perturbation to $H_0$. The
partition function of the reference system,
$\widetilde{\mathcal{Q}}_N^{(0)}$,
is exactly solvable \cite{Baxter}, e.g., using the transfer matrix
$\mathcal{P}\equiv\left(
              \begin{array}{cc}
                e^{J+\alpha_0} & e^{-J} \\
                e^{-J} & e^{J-\alpha_0} \\
              \end{array}
            \right)$ such that
\begin{eqnarray}
\widetilde{\mathcal{Q}}_N^{(0)}
&=&
\sum_{s_1}...\sum_{s_N}e^{-H_0}=
\mathrm{Tr}(\mathcal{P}^N)=\lambda_+^N+\lambda_-^N,
\end{eqnarray}
with
$\lambda_{\pm}=e^J\cosh\alpha_0\pm\sqrt{e^{-2J}+e^{2J}\sinh^2\alpha_0}$
being the eigenvalues of $\mathcal{P}$. Thus,
$\ln\widetilde{\mathcal{Q}}_N^{(0)}\rightarrow N\ln\lambda_+$ for
large $N$. The statistical perturbation theory inspires us to write
$\widetilde{\mathcal{Q}}_N=\widetilde{\mathcal{Q}}_N^{(0)}\sum_{s_1}...\sum_{s_N}
e^{-H_0}e^{-H_1}/\widetilde{\mathcal{Q}}_N^{(0)}
=\widetilde{\mathcal{Q}}_N^{(0)}\langle
e^{-H_1}\rangle\simeq\lambda_+^N[1+\frac{p}{4}\sum_n\langle
s_n\rangle\cos\theta_n+\frac{p^2}{32}\sum_{n,m}\langle
s_ns_m\rangle\cos\theta_n\cos\theta_m]$. Here, we denote
$\theta_n\equiv\varphi_n-\phi$ for short and use $\langle
\cdot\rangle$ to represent the expectation over the reference
system. Due to isotropy, $\langle s_n\rangle$ is independent of its
location (index $n$) and hence  $\sum_n\langle
s_n\rangle\cos\theta_n=\langle s_n\rangle\sum_n\cos\theta_n=0$. We
further calculate that $\sum_{n,m}\langle
s_ns_m\rangle\cos\theta_n\cos\theta_m=\frac{N}{2}(1+2\xi)/(1+e^{4J}\sinh^2\alpha_0)$,
\cite{footnote2}, 
where $\xi\equiv[\ln(\lambda_+/\lambda_-)]^{-1}$
defines the correlation length of the classic Ising chain
\cite{Baxter}. 
Finally, the log-partition function of our model
is
\begin{equation}
\ln\widetilde{\mathcal{Q}}_N\simeq
N\ln\lambda_{+}+\frac{Np^2(1+2\xi)}{64(1+e^{4J}\sinh^2\alpha_0)}+\mathcal{O}(p^3),
\end{equation}
which reduces to Eq. (2) as $J\rightarrow 0$.

Now we rewrite
$\widetilde{\mathcal{H}}_N=-J\sum_ns_ns_{n+1}-\alpha_0z_0-(\alpha_1z_1+\alpha_2z_2)/2$,
with the same notations for $\alpha_i$ and $z_i$, $i=0,1,2$. As has
been demonstrated before, the MLE of $\alpha_1$ and $\alpha_2$ can
be found from the joint Gaussian distribution of $z_1$ and $z_2$,
except now we have to replace $\mu$ by
$\widetilde{\mu}\equiv\frac{1}{16}N(1+2\xi)/(1+e^{4J}\sinh^2\alpha_0)$.
So the MLE of $p$ and $\phi$ are given by
$\widetilde{p}=\widetilde{\mu}^{-1}\sqrt{z_1^2+z_2^2}\xrightarrow{d}
\mathcal{N}(p,\widetilde{\sigma}_p^2)$ and
$\widetilde{\phi}=\arctan(z_2/z_1)\xrightarrow{d} \mathcal{N}(\phi,
\widetilde{\sigma}_{\phi}^2)$. 
Similar to Eq. (5-6), their variances
are $\widetilde{\sigma}_p^2=2/\widetilde{\mu}$ and
$\widetilde{\sigma}_{\phi}^2=\widetilde{\sigma}_{p}^2/p^2=2/
(\widetilde{\mu}p^2)$ \cite{footnote3}.
We plot $\widetilde{\sigma}_{\phi}^2$ as a function of
$\ln(C_0/K_d)$ for different values of $J$ in Fig. 2C. 
Regardless of the receptor
coupling strength,
this error is minimal at $C_0=K_d$ (or $\alpha_0=0$) 
where the correlation length is
$\xi=1/\ln(\coth J)\simeq\frac{1}{2}e^{2J}$ and
$\widetilde{\sigma}_{\phi}^2\simeq32/[Np^2(1+e^{2J})]=\sigma_{\phi}^2/(1+e^{2J})$.

Receptor cooperativity
may help a smaller cell of diameter $\widetilde{L}$ 
achieve the same level of accuracy as a 
larger cell of diameter
$L$ with independent receptors, i.e.,
$\widetilde{\sigma}_{\phi}^2(\widetilde{L})=\sigma_{\phi}^2(L)$.
By our previous scaling assumption, the receptor number of the smaller
cell is $\widetilde{N}=N(\widetilde{L}/L)^{\delta}$. 
If $L^*$ denotes a critical cell length below which spatial
sensing is infeasible with non-cooperative receptors, then adding
cooperativity will push the critical cell size, $\widetilde{L}^*$, 
lower by a factor of $(1+2\xi)^{-1/(2+\delta)}\simeq
L(1+e^{2J})^{-1/(2+\delta)}$. 
This is shown in Fig. 2D where we have plotted $\widetilde{L}^*/L^*$
as a function of $J$ for three values of the scaling factor $\delta$.
As a specific example, we take  $L^* = 8 \mu m$ which corresponds to 
the typical size  of a {\em Dictyostelium} amoeba.
Then, we see that for a cooperativity of $J=0.5$ the 
new critical size becomes $\widetilde{L}^* \sim 4-6
\mu m$, comparable to the size of many bacterial cells. 
It is worth remarking that
although receptor interaction improves the precision of gradient
sensing for $C_0$ close to $K_d$, it enlarges the errors when $C_0$
is far away from $K_d$ (Fig. 2C). In other words, the improved
accuracy near $K_d$ is at the cost of the sensitivity range of
background
concentrations. Such a tradeoff could be a limiting factor for the
introduction of coupling into the spatial sensing
mechanism.

It is commonly believed that prokaryotic cells such as \emph{E.
coli} are too small to perform spatial sensing of chemical
gradients. However, recent experimental observations show that at
least one type of vibrioid bacteria (typical size 2$\times$6 $\mu
m$) are able to  spatially sense gradients 
along distances as short
as 5 $\mu m$ \cite{Thar}. Our results allow for the
possibility that smaller organisms employ a spatial sensing
strategy with the aid of receptor cooperativity. As spatial sensing
is argued to be superior to temporal sensing for fast swimming
bacteria \cite{Dusenbery, Thar}, this possibility is of significant
theoretical interest and remains a challenge for future empirical
studies.

We thank W. Loomis, B. Li, R.J. Williams, J. Wolf, and M. Skoge for
valuable discussions. 
This work was supported by NIH Grant P01 GM078586.


\begin{thebibliography}{Parent}
\bibitem{Parent} C.A. Parent and P.N. Devreotes, Science.
\textbf{284}, 765 (1999).

\bibitem{Haastert} P.J.V. Haastert and P.N. Devreotes, Nat. Rev.
Mol. Cell Biol. \textbf{5}, 626 (2004).

\bibitem{Macnab} R.M. Macnab and D.E. Koshland, Proc. Natl. Acad. Sci. U.S.A.
\textbf{69}, 2509 (1972).

\bibitem{Segall} J.E. Segall \emph{et al.}, Proc. Natl. Acad. Sci. U.S.A.
\textbf{83}, 8987 (1986).

\bibitem{Sourjik} V. Sourjik and H.C. Berg, Proc. Natl. Acad. Sci. U.S.A.
\textbf{99}, 123 (2002).

\bibitem{Bray} D. Bray \emph{et al.}, Nature, \textbf{393},
85 (1998).

\bibitem{Song} L. Song, \emph{et al.}, Eur. J. Cell Biol. \textbf{85}, 981 (2006).

\bibitem{HaastertBJ} P.J.V. Haastert and M. Postma, Biophys. J.
\textbf{93}, 1787 (2007).

\bibitem{Fuller} D. Fuller \emph{et al.}, Proc. Natl. Acad. Sci.
U.S.A. (to be published).

\bibitem{Ueda1} M. Ueda \emph{et al.},
Science, \textbf{294}, 864 (2001).

\bibitem{Ueda2} M. Ueda and T. Shibata, Biophys. J. \textbf{93}, 11 (2007).


\bibitem{Berg1977} H.C. Berg and E.M. Purcell, Biophys. J. \textbf{20}, 193 (1977).

\bibitem{Bialek2005} W. Bialek and S. Setayeshgar, Proc. Natl. Acad. Sci. U.S.A. \textbf{102},
10040 (2005).

\bibitem{Wang1} K. Wang \emph{et al.}, Phys. Rev.
E \textbf{75}, 061905 (2007).

\bibitem{Bialek2008} W. Bialek and S. Setayeshgar, Phys. Rev. Lett. \textbf{100}, 258101 (2008).

\bibitem{EndresPNAS} R.G. Endres and N.S. Wingreen, Proc. Natl. Acad. Sci. U.S.A.
\textbf{105}, 15749 (2008).

\bibitem{EndresPRL} R.G. Endres and N.S. Wingreen, Phys. Rev. Lett. \textbf{103}, 158101
(2009).

\bibitem{Wouter} W.-J. Rappel and H. Levine, Phys. Rev.
Lett. \textbf{100}, 228101 (2008); W.-J. Rappel and H. Levine, Proc.
Natl. Acad. Sci. U.S.A. \textbf{105}, 19270 (2008).

\bibitem{Thar} R. Thar and M. K\"{u}hl, Proc. Natl. Acad. Sci. U.S.A. \textbf{100},
5748 (2003).

\bibitem{footnote1} This assumption is for analytical convenience and can be
relaxed numerically. Our theoretical results work well even
if the receptors are assumed to be uniformly distributed at random
on the cell surface.

\bibitem{Hu} B. Hu \emph{et al.},
Phys. Rev. E \textbf{81}, 031906 (2010).

\bibitem{SteveKay} S.M. Kay, \emph{Fundamentals of Statistical Signal Processing: Estimation Theory}
(Prentice Hall PTR, Upper Saddle River, NJ, 1993), Vol. 1, Chap. 3.

\bibitem{Lauffenburger} D.A. Lauffenburger and J.J. Linderman, \emph{Receptors: Models for Binding,
Trafficking, and Signaling} (Oxford University Press, New York,
1993), Chap. 4.

\bibitem{Shi} Y. Shi and T. Duke, Phys. Rev. E \textbf{58}, 6399 (1998).

\bibitem{Mello} B.A. Mello and Y. Tu, Proc. Natl. Acad. Sci. U.S.A.
\textbf{100}, 8223 (2003); 
B.A. Mello and Y. Tu, \emph{ibid}. 
\textbf{102}, 17354 (2005);
B.A. Mello \emph{et al.}, Biophys.
J. \textbf{87}, 1578 (2004). 

\bibitem{Wingreen} J.E. Keymer \emph{et al.}, Proc. Natl.
Acad. Sci. U.S.A. \textbf{103}, 1786 (2006); M.L. Skoge, R.G.
Endres, and N.S. Wingreen, Biophys. J. \textbf{90}, 4317 (2006).

\bibitem{Baxter} R.J. Baxter, \emph{Exactly Solved Models in Statistical
Mechanics} (Academic, London, 1982), Chap. 2.


\bibitem{footnote2} For the classic Ising chain, the spin-spin correlation is $\langle
s_ns_m\rangle=\cos^22\omega+\gamma^{|n-m|}\sin^22\omega$, where
$\gamma=\lambda_-/\lambda_+$ and $\omega$ is defined by the equation
$\cot{2\omega}=e^{2J}\sinh\alpha_0$ for $0<\omega<\pi/2$
\cite{Baxter}. Thanks to
$\sum_{n}\sum_m\cos^22\omega\cos\theta_n\cos\theta_m=\cos^22\omega\sum_{n}\cos\theta_n\sum_m\cos\theta_m=0$,
we only need calculate
$\sum_{n}\sum_m\cos\theta_n\cos\theta_m\gamma^{|n-m|}=\sum_n\cos\theta_n\sum_m\cos[(\theta_m-\theta_n)+\theta_n]\gamma^{|n-m|}=
\sum_n\cos^2\theta_n\sum_m\cos(\theta_m-\theta_n)\gamma^{|n-m|}-\sum_n\cos\theta_n\sin\theta_n\sum_m\sin(\theta_m-\theta_n)\gamma^{|n-m|}$.
The second term vanishes since we have
$\sum_m\sin(\theta_m-\theta_n)\gamma^{|n-m|}
=\sum_{j=-N/2}^{N/2}\sin(2\pi j/N)\gamma^{|j|}=0$, while the first
term above is identical to
$\sum_n\cos^2\theta_n\sum_{j=-N/2}^{N/2}\cos(2\pi
j/N)\gamma^{|j|}=\sum_n\cos^2\theta_n\left[1+2\sum_{j=1}^{N/2}\cos(2\pi
j/N)\gamma^j\right]$. For large $N$, we have that
$\sum_n\cos^2\theta_n\simeq N/2$ and $\sum_{j=1}^{N/2}\cos(2\pi
j/N)\gamma^j\simeq\frac{N}{2\pi}\int_0^{\pi}\cos(x)\exp\left(\frac{xN}{2\pi}\ln\gamma\right)dx
\xrightarrow{N\rightarrow\infty}-1/\ln\gamma\equiv\xi$. Thus,
$\sum_{n,m}\langle
s_ns_m\rangle\cos\theta_n\cos\theta_m=\frac{N}{2}(1+2\xi)\sin^22\omega=\frac{N}{2}(1+2\xi)/(\cot^22\omega+1)
=\frac{N}{2}(1+2\xi)/(1+e^{4J}\sinh^2\alpha_0)$.

\bibitem{footnote3} 
As long as we are not near any
phase transition point, the measurement decorrelation 
time will remain dominated by the processes of  diffusion and
binding/unbinding of ligand molecules \cite{Bialek2008}. 
Thus, 
averaging signals over $\mathcal{T}$ will give
$\widetilde{\sigma}_{p,\mathcal{T}}^2\simeq4\tau_{\mathrm{rec}}(1+\eta)/(\widetilde{\mu}\mathcal{T})$
and
$\widetilde{\sigma}_{\phi,\mathcal{T}}^2=\widetilde{\sigma}_{p,\mathcal{T}}^2/p^2$.

\bibitem{Dusenbery} D.B. Dusenbery,  Biophys. J. \textbf{74}, 2272
(1998).

\end{thebibliography}
\end{document}